\documentclass[aps,twocolumn,floatfix,showpacs,amsmath,amsfonts]{revtex4}

\usepackage{graphicx}
\usepackage{bm}
\usepackage{array}
\usepackage{hyperref}

\begin{document}

\title{Signatures of the classical transition state in atomic quantum spectra}

\author{Holger Cartarius}
\email{Holger.Cartarius@itp1.uni-stuttgart.de}
\author{J\"org Main}
\author{G\"unter Wunner}
\affiliation{Institut f\"ur Theoretische Physik 1, Universit\"at Stuttgart,
  70550 Stuttgart, Germany}
\date{\today}

\begin{abstract}
We perform quantum mechanically exact calculations of resonances in the
spectrum of the hydrogen atom in crossed external fields and establish a close
connection between the classical transition state in phase space and features
in the quantum spectrum. By varying the external field strengths, structures
are revealed which are surprisingly similar to the quantized energy levels of
the classical electron motion in the vicinity of the saddle point obtained with
an approximation of the potential. The results give clear evidence for 
signatures of the transition state in quantum spectra.
\end{abstract}

\pacs{32.60.+i, 32.80.Fb, 82.20.Db}

\maketitle

\section{Introduction}
\label{sec:intro}

Transition state theory, which has its original applications in the theory of
chemical reactions (see, e.g., \cite{Keck67,Eyr31,Eva35}), can also be applied
to many dynamical systems which evolve from an initial (``reactants'') to a
final state (``products''). 
The key concept is based on trajectories in the classical phase space of
a dynamical system. In particular, regions of the phase space which identify
the reactant side and those which represent the product side are introduced.
Of special interest are trajectories that connect the reactant side with the
product side, because they describe the ``reaction.''  The theory postulates the
existence of a minimal set of states which is passed by all of these
``reactive'' trajectories, and this set is called the transition state.
The transition state has no intersection with a ``nonreactive'' trajectory,
and therefore describes a \emph{boundary} between reactants and
products in the phase space.

\begin{figure}[tb]
  \centering
  \includegraphics[width=\columnwidth]{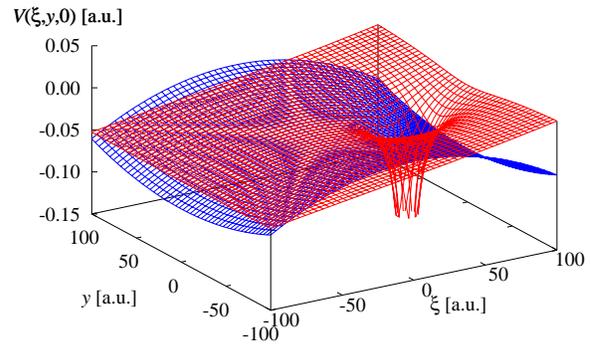}
  \caption{\label{fig:saddle}(Color online) Potential of the hydrogen atom
    in an external electric field in the ($\xi$,$y$)-plane (cf.\ Eq.\
    \eqref{eq:pot_exact}) with $\xi = x-x_s$ denoted by the solid
    red grid. Reactive trajectories must pass the Stark saddle point at
    $(\xi=0,y=0)$ in whose vicinity the transition state is located. The
    dashed blue grid describes a quadratic approximation around the saddle
    point (cf.\ Eq.\ \eqref{eq:potential_quadratic}).}
\end{figure}

Among the large number of applications of the transition state theory,
it has also proven to be important for understanding the
ionization mechanism of atoms. In the case of the hydrogen atom in an external
electric field the ``reaction'' corresponds to the ionization. The potential
of the system is plotted in Fig.\ \ref{fig:saddle} (solid red grid), in which
the Stark saddle point is clearly visible.
Classical trajectories describing an ionization must pass the saddle point,
and therefore the transition state must be located in its vicinity.
An early work by Clark et al.\ \cite{Cla85} can be regarded as a first step
in discussing the ionization mechanism of the hydrogen atom in crossed
electric and magnetic fields in this context. They used a quadratic expansion of
the potential around the Stark saddle point (see Fig.\ \ref{fig:saddle},
dashed blue grid) to calculate quasi-bound states confined
to the vicinity of the saddle. Further progress was achieved
by Jaff\'e et al.\ \cite{Jaf00,Jaf99} who discussed transition state
theory for systems without time-reversal symmetry, and found the transition
state of the planar (i.e., two-dimensional) hydrogen atom in crossed external
electric and magnetic fields. With the help of methods from nonlinear dynamics
Jaff\'e et al.\ \cite{Jaf00,Jaf99} discussed the chaotic ionization
mechanism, and were able to explain the existence of electrons ionizing promptly
and electrons ionizing with delay after repeated encounters with the core
region. The problem that the practical applicability of transition state
theory was restricted to low-dimensional systems was overcome by an
algorithmic procedure which allows identifying the transition state in
higher-dimensional systems \cite{Wig01,Uzer02,Waa08}, and is based on a
normal-form representation of a power series expansion of the Hamiltonian. The
quadratic approximation used by Clark et al.\ \cite{Cla85} coincides with the
lowest order in the normal form expansion of the Hamiltonian introduced by
Uzer et al.\ \cite{Uzer02}. 

Even though the transition state for the hydrogen atom in crossed fields had
been identified, the question of whether or not the classical trajectories in
its vicinity leave a signature in the exact quantum spectrum remained
unanswered since no such quantum spectra were available. We have succeeded in
calculating these spectra, and it is the purpose of this paper to establish the
close connection between the classical motion of an electron confined near the
transition state and exact quantum resonances. To calculate the electron
trajectories near the
transition state we adhere to the simplest approximation and use the
power series expansion of the potential around
the Stark saddle point up to second order introduced by Clark et al.\
\cite{Cla85}. Higher orders would rapidly lead to a drastic increase of the
computational effort, and, as will be seen from the results, the agreement of
the quantized energy levels of the electron's motion with resonances in the
quantum mechanically exact spectrum is already extremely good for the simple
expansion. Thus, the expansion is completely sufficient to reveal the
signatures of the transition state in the quantum spectrum.

The crossed-fields hydrogen atom is an ideal candidate to search for
the connection between the classical transition state and its counterpart
in a quantum system because the lowest-order approximations to the transition
state as well as the quantum states are accessible with reasonable numerical
effort. Chemical reactions with their large number of degrees of freedom are
far more complicated and do not allow for the insight we gain in the case of
the hydrogen atom, where only three degrees of freedom exist.

The paper is organized as follows. The system and the approximation around
the saddle point are introduced in Sec.\ \ref{sec:Hamiltonian_resonances}.
In Sec.\ \ref{sec:discussion}  we present the results for the resonances.
The calculation is performed for different external fields, which are varied
such that lines in the two-dimensional space of the field strengths are
traversed. This procedure will uncover close relations in the energies
calculated with both methods. Finally, conclusions are drawn in Sec.\
\ref{sec:conclusions}.

\section{Hamiltonian and resonances in the vicinity of the Stark
  saddle point}
\label{sec:Hamiltonian_resonances}

The Hamiltonian of the hydrogen atom in crossed electric and magnetic fields
is given, in atomic units, by
\begin{equation}
  H = \frac{1}{2} \left ( \bm{p} + \bm{A} \right )^2 - \frac{1}{r} 
  + f x \; ,
  \label{eq:Hamiltonian_exact}
\end{equation} 
where the electric field $f$ is orientated along the $x$-axis and $\bm{A}$
represents the vector potential. The magnetic field is supposed to be
oriented along the $z$-axis. In this case, usually the symmetric gauge
$\bm{A} = -1/2 (\bm{r}\times \gamma\bm{e}_z)$ with the magnetic flux density
$\gamma$ is used, whereas, for the approximated states discussed in
this paper the application of  the Landau gauge $\bm{A} = (-\gamma y,0,0)$ is
more suitable. Besides the energy the parity with respect to the ($z=0$)-plane
is a good quantum number, which opens the possibility to consider states with
even and odd $z$-parity separately.

\subsection{Quadratic approximation}

The transition state is located in the vicinity of the Stark saddle point,
where the net electric force vanishes. For the potential caused by the
external electric field and the nucleus, 
  \begin{equation}
    V_f = -\frac{1}{r} + f x \; ,
    \label{eq:pot_exact}
  \end{equation}
the saddle point is located on the $x$-axis at $x_s = -1/\sqrt{f}$,
and the saddle point energy has the value $V_f(\bm{r}_s) = -2\sqrt{f}$. The
simplest approximation of the potential $V_f$ which contains the
structure of the saddle, is a quadratic expansion.
This expansion leads to unbound solutions called ``Quasi-Penning'' resonances
by Clark et al.\ \cite{Cla85} due to a formal similarity with the motion in
a Penning trap. For the readers' convenience, we recapitulate the essential
results. The expansion of the potential up to second order terms reads
\begin{equation}
  V_f(\bm{r}) = -2\sqrt{f} - \sqrt{f}^3 \xi^2 + \frac{1}{2} \sqrt{f}^3
  \left (y^2 + z^2 \right ) + \mathrm{O}((\bm{r}-\bm{r}_s)^3)\; ,
  \label{eq:potential_quadratic}
\end{equation}
where $\xi = x-x_s$. The saddle structure of the potential has been visualized
in Fig.\ \ref{fig:saddle}, where the potential \eqref{eq:potential_quadratic}
is plotted in the ($\xi$,$y$)-plane. The comparison with the exact electric
potential \eqref{eq:pot_exact} shows clearly that the approximation is only
valid close to the saddle point and can only describe states in its vicinity
correctly.

With the gauge $\bm{A} = (-\gamma y,0,0)$ the Hamiltonian in the vicinity of
$\bm{r}_s$ reads
\begin{multline}
  H = \frac{1}{2} \left ( p_\xi^2 + p_y^2 + p_z^2 \right )
  - \gamma y p_\xi + \frac{1}{2}\gamma^2 y^2 \\
  - 2\sqrt{f} + \frac{1}{2}\sqrt{f}^3\left ( y^2 + z^2 
    - 2 \xi^2\right ) \; ,
\end{multline}
and the eigenvalues of the quadratic potential yield the energy levels
\begin{multline}
  E_{n_z,n_1,n_2} = -2\sqrt{f} + \omega_z(n_z+\frac{1}{2}) \\
  + \omega_1(n_1+\frac{1}{2}) + \omega_2(n_2+\frac{1}{2}) \; ,
\end{multline}
where $\omega_z$ represents the frequency of the $z$-motion and is
given by $\omega_z = f^{3/4}$. The separation of the coupled equations in the
$\xi$- and $y$-directions with the help of an adequate canonical
transformation \cite{Uzer02}, leads to one real oscillation frequency,
\begin{equation}
  \omega_1 = \left \{\frac{1}{2}\left(\gamma^2 - \sqrt{f}^3 +
      \sqrt{\left ( \gamma^2 - \sqrt{f}^3 \right )^2 + 8f^3} \right )
  \right \}^{1/2} \; , 
\end{equation}
and one imaginary decay rate,
\begin{equation}
  \omega_2 = \mathrm{i} \left \{\frac{1}{2}\left( \sqrt{f}^3 - \gamma^2
      +\sqrt{\left ( \gamma^2 - \sqrt{f}^3 \right )^2 + 8f^3} \right ) 
  \right \}^{1/2} \; ,
\end{equation}
which describes the resonance character of the eigenstates. Thus, the
resonance energies (real part of the complex eigenvalues) are completely
determined by the two quantum numbers $n_z$, $n_1$ and read
\begin{equation}
  \mathrm{Re}(E)_{n_z,n_1} = -2\sqrt{f} + \omega_z(n_z+\frac{1}{2})
  + \omega_1(n_1+\frac{1}{2}) \; .
  \label{eq:qp_energies}
\end{equation}

\subsection{Exact quantum calculations}

The aim of this paper is to present evidence for the occurrence of quantum
resonances corresponding to the classical transition state. For this purpose
the approximate energy values \eqref{eq:qp_energies} are compared with the
results of exact quantum calculations. To perform the quantum calculations the
Hamiltonian \eqref{eq:Hamiltonian_exact} is rewritten in dilated semiparabolic
coordinates
\begin{equation}
  \mu = \frac{1}{b}\sqrt{r+z}\; ,\quad \nu = \frac{1}{b}\sqrt{r-z}\; , \quad
  \varphi = \frac{y}{x}\; ,
\end{equation}
where $b$ is a convergence parameter. The transformed Schr\"odinger equation
reads \cite{Mai94}
\begin{multline}
  \bigg \{ \Delta_\mu + \Delta_\nu - \left ( \mu^2 + \nu^2 \right ) 
  + b^4 \gamma \left ( \mu^2 + \nu^2 \right ) \mathrm{i}\frac{\partial}{\partial
    \varphi} \\ 
  - \frac{1}{4} b^8 \gamma^2 \mu^2\nu^2 \left ( \mu^2 + \nu^2 \right ) 
  - 2 b^6 f \mu\nu \left ( \mu^2 + \nu^2 \right ) \cos \varphi \bigg \} \psi\\ =
  \left \{ -4b^2 + \lambda \left ( \mu^2 + \nu^2 \right ) \right \} \psi 
  \label{eq:Schroedinger_transformed}
\end{multline}
with 
\begin{equation}
  \Delta_\varrho = \frac{1}{\varrho} \frac{\partial}{\partial\varrho} \varrho
  \frac{\partial}{\partial\varrho} + \frac{1}{\varrho^2} 
  \frac{\partial^2}{\partial\varphi^2} \;, \qquad \varrho \in \left \{ \mu,
    \nu \right \} \; ,
\end{equation}
and the generalized eigenvalues $\lambda = -(1+2b^4 E)$, which are related to
the energies $E$ of the quantum states. We use a matrix representation of
the full Schr\"odinger equation \eqref{eq:Schroedinger_transformed} in a
complete discrete set of basis functions of the form
\begin{equation}
  | N_\mu, N_\nu, m \rangle = | N_\mu, m \rangle \otimes | N_\mu, m \rangle \;,
\end{equation}
where $|N_\varrho m\rangle$ are the eigenstates of the two-dimensional 
harmonic oscillator. Resonances are uncovered with the complex rotation method
\cite{Rei82,Del91}. Here, the coordinates $\bm{r}$ of the system are replaced
with complex rotated ones $\bm{r}\mathrm{e}^{\mathrm{i} \vartheta}$, which leads
to a complex symmetric matrix representation of the Hamiltonian. The complex
rotation can be inserted in the transformed Schr\"odinger Eq.\
\eqref{eq:Schroedinger_transformed} with the help of the complex extended
convergence parameter $b$
\begin{equation}
  b = |b| \mathrm{e}^{\mathrm{i}\vartheta/2}\; .
\end{equation}
Resonances appear as complex eigenvalues $E$, where the real part of $E$
represents the energy and the imaginary part is related to the width
$\Gamma = -2\mathrm{Im}(E)$. 

With this method, the resonances of the hydrogen atom can be calculated with a
high precision. It must be mentioned, however, that  techniques for the matrix
diagonalization used so far may fail in the energy region of interest.
Former approaches to calculating the exact quantum resonances were mainly based
on complex extensions of very efficient diagonalization algorithms for real
symmetric matrices (see, e.g., Refs.\ \cite{Mai92,Mai94}) which require an
orthogonality relation between all eigenstates. Of course, an orthogonality
relation is not assured for non-Hermitian Hamiltonians and is, in particular,
violated at so-called exceptional points, whose existence in the
relevant part of the spectrum has been proven recently \cite{Car07a}.
Exceptional points are branch point singularities of the resonances at which
both the complex energies and the wave functions of two states become
identical. 
In our calculations, the diagonalization was performed by applying the
implicitly restarted Arnoldi method \cite{Leh98}, which solves large scale
sparse eigenvalue problems efficiently even for non-Hermitian Hamiltonians. 
The matrices were built up with a basis of about 10,000 states.

\section{Results and discussion}
\label{sec:discussion}

To decide whether or not there is a connection between the classical motion
around the saddle point and features in the exact quantum spectra, we compare
the energy eigenvalues \eqref{eq:qp_energies} with the exact quantum
resonances. A coincidental agreement between two energies of both
approaches for a \emph{single} parameter value is always possible and does not
answer our question. However, a connection between the approximation
\eqref{eq:qp_energies} and the quantum resonances is certainly present if
the two results agree in a \emph{larger region} of the parameter space.
To check the connection we perform the calculations on lines $(f-f_0)/\gamma 
= \mathrm{const}$ in the two-dimensional parameter space spanned by the two
field strengths, i.e., we introduce a parameter $\alpha$ and consider the field
strengths $f(\alpha)$ and $\gamma(\alpha)$ as functions of $\alpha$.

\begin{figure}[tb]
  \centering
  \includegraphics[width=\columnwidth]{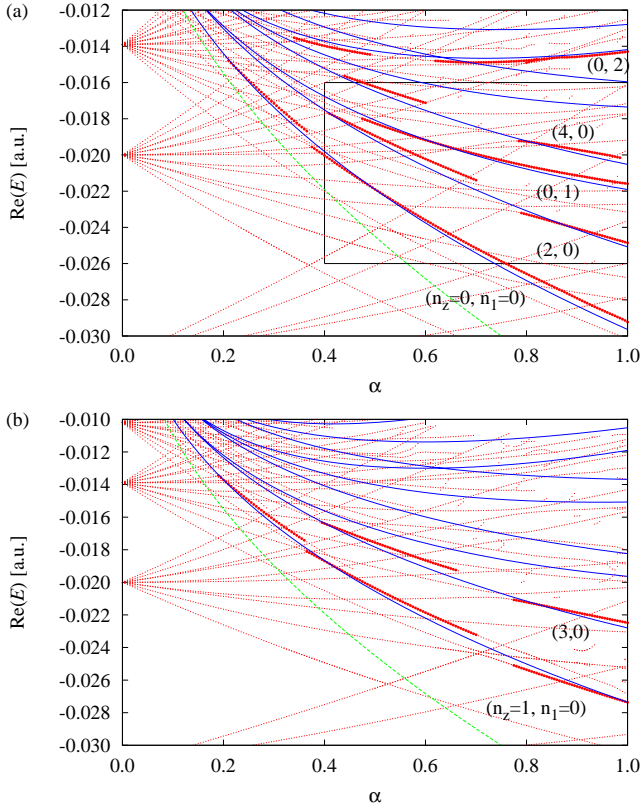}
  \caption{\label{fig:comparison_1}(Color online) (a) Comparison of the
    quantized energies of the electron motion near the transition state (solid
    blue lines) with the exact quantum energies (red points) on the line
    $\gamma = 0.008\times \alpha$, $f = 0.0003 \times \alpha$ for $0 < \alpha
    < 1$ as defined in Eqs.\ \eqref{eq:line_1a}--\eqref{eq:line_1c} in the
    ($\gamma$,$f$)-space for even $z$-parity. Only exact quantum energies
    with $|\mathrm{Im}(E)| < 0.0006$ are shown.
    The exact resonances which trace the approximate energies
    \eqref{eq:qp_energies} are shown by a larger line strength.
    The lowest (dashed green) line represents the saddle point energy.
    The black frame marks the region magnified in Fig.\ 
    \ref{fig:comparison_1_avoided}. 
    (b) Same situation for states with odd $z$-parity.}
\end{figure}

An example for a comparison is presented in Fig.\ \ref{fig:comparison_1}(a)
for states with even $z$-parity.
Here, the resonances are calculated for different external field strengths
on the line defined by
\begin{subequations}
  \begin{align}
    \gamma &= 0.008\times \alpha \; , \label{eq:line_1a} \\
    f &= 0.0003 \times \alpha\; ,  \\
     0 &< \alpha < 1\; , \label{eq:line_1c}
  \end{align}
\end{subequations}
i.e., the ratio $\gamma/f$ of the two field strengths is always
constant. The solid blue lines represent the energy eigenvalues
\eqref{eq:qp_energies} of the quadratic approximation around the saddle point,
and the red points are the numerically exact resonances of the Hamiltonian
\eqref{eq:Hamiltonian_exact}. The results show that for the lowest quantum
numbers $n_z$, $n_1$ of the approximation there are very good agreements
between both approaches on a large range of parameters $\alpha$. The
``Quasi-Penning'' resonances are traced by the quantum mechanically exact
states when $\alpha$ is changed. Wherever this behavior is present, it is
highlighted by a larger line thickness for the exact resonances.

Of course, the effect does also appear for odd $z$-parity, which is shown
separately in Fig.\ \ref{fig:comparison_1}(b) for the same physical parameters
as in Fig.\ \ref{fig:comparison_1}(a). As one can see, the agreement of the
resonances is again very good in large regions of the parameter space.

\begin{figure}[tb]
  \centering
  \includegraphics[width=\columnwidth]{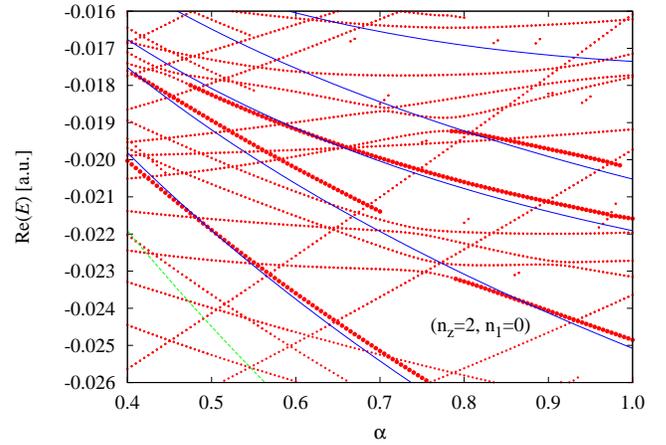}
  \caption{\label{fig:comparison_1_avoided}(Color online) Same situation as
    depicted in Fig.\ \ref{fig:comparison_1}. The region around the
    avoided level crossings near the state with $n_z = 2$, $n_1 = 0$
    is magnified.}
\end{figure}

There are, however, parameter values $\alpha$ for which a different
behavior of the energy values exists. For low field strengths the Coulomb
potential becomes too dominant, and the expansion of the potential in
the vicinity of the Stark saddle point no longer leads to reasonable
solutions, as can be observed in Fig.\ \ref{fig:comparison_1}.
Furthermore, avoided level crossings of the numerically exact
resonances have no counterpart in the approximative states. Evidently,
couplings of states, which are the origin of avoided level crossings, are
not included in the simple model. Thus, the deviations in these regions are not
surprising. Fig.\ \ref{fig:comparison_1_avoided} shows an example of a region
with avoided level crossings of quantum mechanically exact resonances.
It can be clearly seen that far from the crossing the agreement between
the numerically exact energies and the energy value corresponding to $n_z = 2$,
$n_1 = 0$ in Eq.\ \eqref{eq:qp_energies} is very good, whereas at the
avoided crossings themselves the approximative solutions cannot reproduce the
results of the full Hamiltonian \eqref{eq:Hamiltonian_exact}.

\begin{figure}[tb]
  \centering
  \includegraphics[width=\columnwidth]{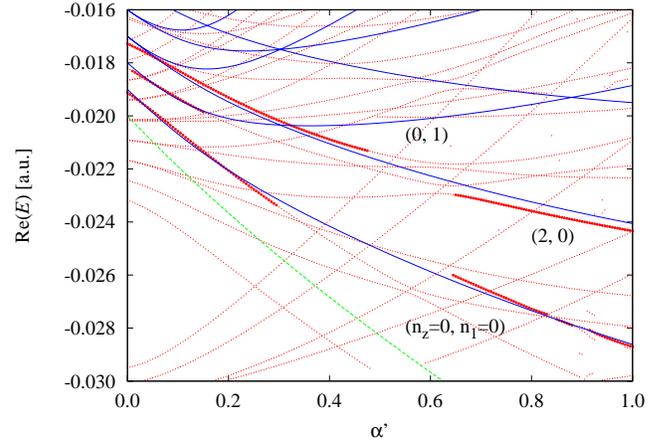}
  \caption{\label{fig:comparison_2}(Color online) Same comparison of the
    quantized energies of the electron motion near the transition state (solid 
    blue lines) with exact quantum energies (red points) as in Fig.\
    \ref{fig:comparison_1} but on the line $\gamma = 0.01\times
    \alpha'$, $f = 0.0001 + 0.0002\times \alpha'$ for $0 < \alpha' < 1$ defined
    in Eqs.\ \eqref{eq:line_2a}--\eqref{eq:line_2c}.
    Only exact quantum energies with $|\mathrm{Im}(E)| < 0.0002$ are shown.
    The saddle point energy is marked by the lowest (dashed green) line.}
\end{figure}

The calculations shown in Fig.\ \ref{fig:comparison_2}
are performed on the line
\begin{subequations}
  \begin{align}
    \gamma &= 0.01\times \alpha' \; , \label{eq:line_2a} \\
    f &= 0.0001 + 0.0002\times \alpha'\; ,  \\
    0 &< \alpha' < 1\; , \label{eq:line_2c}
  \end{align}
\end{subequations}
i.e., the ratio of the field strengths is no longer constant. For $\alpha' = 0$
there is still an electric field present but no magnetic field. The
important property of this line is the existence of the saddle point for all
values of $\alpha'$ drawn in Fig.\ \ref{fig:comparison_2}. Again, there are
exact quantum resonances which behave like the approximate solutions around
the Stark saddle point and lead to a good agreement between both methods
on a wide range of parameter values. Also near $\alpha' = 0$,
where a pure Stark effect is present, the classical electron motion near
the transition state describes the quantum resonances very well.

The calculations performed here reveal that a close connection between
the two approaches exists. It appears for the seven pairs of quantum numbers
of the ``Quasi Penning'' resonances listed in Table \ref{tab:quantum_numbers}.
\begin{table}
    \caption{\label{tab:quantum_numbers}Quantum numbers $n_z$ and $n_1$ of
      the ``Quasi-Penning'' resonances closely related with exact quantum
      states.}
    \begin{ruledtabular}
      \begin{tabular}{>{\centering\arraybackslash}p{.25\columnwidth}
          >{\centering\arraybackslash}p{.25\columnwidth} 
          | >{\centering\arraybackslash}p{.25\columnwidth}
          >{\centering\arraybackslash}p{.25\columnwidth}}
        $n_z$ & $n_1$ & $n_z$ & $n_1$ \\
        \hline
        0 & 0 & 4 & 0 \\
        1 & 0 & 0 & 1 \\
        2 & 0 & 0 & 2 \\
        3 & 0 & & \\
      \end{tabular}
    \end{ruledtabular}
\end{table}
This rather high number of concordant resonances shows that, indeed, there
is a signature of the classical transition state in the exact quantum spectrum.

\section{Conclusion}
\label{sec:conclusions}

In summary we have demonstrated that there is clear evidence for the
correspondence of classical motion of electrons confined to the vicinity
of the Stark saddle point and resonances in the exact quantum spectra.
The agreement of the energies in the classical transition state approach
and the quantum calculations is very good. This appears surprising because the
simplest expansion of the potential, only up to second order, was used, i.e.,
the approximation is only valid in a small region very close to the saddle
point. This clearly can be understood as a sign for an eminent impact of the
transition state on the exact quantum spectrum, and demonstrates the importance
of the transition state for the ionization mechanism.

It must be noted, however, that the excellent agreement between both methods
exists only for the real parts of the complex energy eigenvalues. The simple
approximation used in the expansion of the potential is not capable of
reproducing the correct decay rates or imaginary parts.

As an outlook, it seems worthwhile to extend the calculations to the normal
form procedure developed by Uzer et al.\ \cite{Uzer02} and to check whether
higher orders can lead to an even better description of the quantum resonances
by classical electronic motion. As can be expected, only the lowest
``Quasi-Penning'' resonances turn out to have a counterpart in the exact
quantum resonances. It will be interesting to investigate which are the highest
quantized levels of the transition state theory with a correspondence in the
quantum spectrum.

\begin{acknowledgments}
  This work was supported by Deutsche Forschungsgemeinschaft. H.C. 
  is grateful for support from the Landesgraduiertenf\"orderung of
  the Land Baden-W\"urttemberg.
\end{acknowledgments}

\end{document}